**Title.** Current status of antihistamines repurposing for infectious diseases

Bruno L. Travi[a]

[a]Internal Medicine-Division of Infectious Diseases, University of Texas Medical Branch, Galveston, Texas.

## Abstract

**Objectives.** The principal objectives of this review were to gather information on the potential role of antihistamines as anti-infective agents and to identify the gaps in research that have impaired its applicability in human health.

**Methods.** The literature search was carried using MEDLINE, PubMed and Google Scholar and included all articles in English from January 1990 to May 2022.

**Results.** The literature search identified 12 antihistamines with activity against different pathogens. Eight molecules were second-generation antihistamines with intrinsically lower tendency to cross the blood brain barrier thereby with reduced side effects. Only five antihistamines had *in vivo* evaluations in rodents while one study utilized a wax moth model to determine astemizole anti-*Cryptococcus sp.* activity combined with fluconazole. *In vitro* studies showed that clemastine was active against *Plasmodium, Leishmania,* and *Trypanosoma*, while terfenadine suppressed *Candida* spp. and *Staphylococcus aureus* growth. *In vitro* assays found that SARS-coV-2 was inhibited by doxepin, azelastine, desloratadine, and clemastine. Different antihistamines inhibited Ebola virus (diphenhydramine, chlorcyclizine), Hepatitis C virus (chlorcyclizine), and Influenza virus (carbinoxamine, chlorpheniramine). Generally, *in vitro* activity ($IC_{50}$) of antihistamines was in the low to sub-µM range, except for *Staphylococcus epidermidis* (loratadine MIC=50 µM) and SARS-coV-2 (desloratadine 70% inhibition at 20 µM).

**Conclusion.** Many antihistamine drugs showed potential to progress to clinical trials based on *in vitro* data and availability of toxicological and pharmacological data. However, the overall lack of systematic preclinical trials has hampered the advance of repurposed antihistamines for off label evaluation. The low interest of pharmaceutical companies has to be counterbalanced through collaborations between research groups, granting agencies and government to support the needed clinical trials.

**Author Address**:
Bruno L. Travi, DVM, PhD
University of Texas Medical Branch (UTMB)
301 University Boulevard
4.318 Marvin Graves Building
Galveston, Texas  77555-0435
Phone: (409) 747-4318
brltravi@utmb.edu

**Key words**: antihistamines-repurposing-anti-infective-pathogens-chemical library

## 1. Introduction

The repurposing of FDA-approved drugs for off-label utilization is considered to be a more cost-effective and less risky approach than de novo drug discovery for infectious diseases. Repurposing candidate drugs has the advantage of selecting compounds that have gone through several stages of clinical development including pharmacokinetic profiles and safety data that allow its rapid utilization in patients [1-3]. Several repurposed drugs in clinical use were initially prescribed to treat specific illnesses but unanticipated off-target effects identified new therapeutic indications. Some outstanding examples are, sildenafil, thalidomide, bupropion and minoxidil [3].

Drug repurposing is currently based on the systematic screening of large chemical libraries. This review compiles the existing information on antihistamine drugs with potential antipathogen activity identified through the screening of libraries containing FDA-approved drugs. Antihistamine drugs were developed for allergy-related pathologies and are widely used in the clinic. First generation antihistamines cross the blood brain barrier and tend to have important side effects such as sedation, decreased cognitive function, and orthostatic hypotension. However, 2$^{nd}$ and 3$^{rd}$ generation antihistamines do not cross the blood brain barrier and tend to produce less side effects [4].

The activity of antihistamines is exerted through their binding to one of four known histamine receptors (HRs). However, H1R and H2R are the targets of virtually all antihistamine drugs in clinical use. These two receptors are expressed in nerve cells, airway and vascular smooth muscles, hepatocytes, as well as immune cells (neutrophils, eosinophils, monocytes, DC, T and B cells) [5]. Antihistamines act as inverse agonists, meaning that they bind to the inactive form of the histamine receptors blocking the activation that leads to cell signaling, gene transcription and specific cellular responses [5].

This review focuses exclusively on the anti-infective capacity of antihistamines through their interaction with distinct pathogen targets ranging from viruses to metazoan parasites. Drug screening of large chemical libraries and utilization of *in vitro* studies strongly indicated that multiple antihistamines interact with undescribed pathogen targets. So far, not all the mechanisms of action have been fully elucidated, which makes drug optimization through medicinal chemistry more difficult. Nevertheless, many of the currently FDA-approved antihistamines have potential as anti-infective agents, used alone or more likely in combination with other drugs.

## 2. Methods

The literature search was performed using PubMed, MEDLINE and Google Scholar online databases. Article selection included all publications in English from January 1990 to May 2022. The MEDLINE search strategy used the following string: (*antihistamine AND virus infection*) OR (*antihistamine AND bacterial infection*) OR (*antihistamine AND parasite infection*) OR (*antihistamine AND fungal infection*). Search in PuMed used the following strategy: ((antihistamine) AND (viral infection)) AND (repurposing); ((antihistamine) AND (bacterial infection)) AND (repurposing); ((antihistamine) AND (fungal infection)) AND (repurposing); ((antihistamine) AND (parasite infection)) AND (repurposing). Search in Google Scholar used the following key words: antihistamine virus infection, antihistamine bacterial infection, antihistamine parasite infection, antihistamine fungal infection. All the articles that generated *in vitro* or *in vivo* data on antihistamine(s) activity against pathogen were evaluated.

### 3. Activity against parasites

**3.1. Plasmodium spp.**

*Epidemiological importance.* Despite sustained control efforts supported by international agencies and local governments, WHO estimates that malaria endemicity accounts for 229 million annual cases and more than 400,000 annual deaths, principally in African countries [6]. In fact, the Africa reports 94% of cases and deaths, affecting mainly children under 5 years of age. Cerebral malaria, for which there is no specific treatment other than supporting therapy, is a severe and poorly understood pathology that seems to be driven by a dysregulated and inflammatory host response to the parasite [7](Dunst, 2017).

*Current treatment and mechanism of action.* For uncomplicated *P. falciparum* malaria, which represents the larger proportion of cases, different drug combinations of artesunate with amodiaquine, mefloquine or sulfadoxine-pyrimethamine and other drugs (arthemeter or artemisinin, etc.) are currently used [8]. *P. vivax* infections have to be treated with chloroquine together with primaquine or tafenoquine to eliminate the liver dormant stage (hypnozoite) responsible for recurrent disease [9].

Chloroquine and its derivatives (amodiaquine) interfere with the parasite's heme detoxification capacity, promoting its lethal accumulation in the intra-erythrocyte stage of *Plasmodium* sp. [10]. Like chloroquines, artemisinin and its analogues inhibit polymerization of heme, and also alter membrane transport activity of the parasite inhibiting nutrient intake [11]. The most probable mechanism of action of primaquine derives from its transformation to hydroxyl metabolites by the CYP-2D6 liver enzyme generating parasiticidal reactive oxygen species. The mode of action of tafenoquine is less clear but could be similar, in part, to primaquine [9]. Importantly, parasite drug resistance is a lingering problem requiring the identification of new molecules that used alone or in combination with existing drugs could improve treatment efficacy, thereby decreasing morbidity, mortality, and transmission [11, 12].

*Antihistamines and mechanism of action.* Several studies have focused on astemizole, a 2$^{nd}$ generation antihistamine that inhibits H1R activation and has demonstrated to have antimalarial activity. Initial studies by Chong et al. using cultured *P. falciparum* to screen 1,937 FDA-approved drugs from the Johns Hopkins Clinical Compound Library (JHCCL) identified astemizole as a potential antimalarial drug [13] (Supp. Table 1). Cardiac arrhythmias may occur because of astemizole overdose or when it is co-administered with drugs that block its metabolism through cytochrome P450 3A5 (CYP 3A4) [13]. However, data from multiple countries showed that the fatality rate is approximately 8 per million doses sold [14]. Astemizole and its metabolite desmethylastemizole showed anti-*Plasmodium* activity at nanomolar concentrations varying between 57 and 117 nM depending on parasite strain. Additionally, many astemizole analogues were synthesized with the purpose of preserving or increasing the antimalarial activity while reducing the hERG-blocking capacity, which is infrequently associated with cardiac toxicity [15, 16].

The increasing multi-drug resistance found in *P. falciparum* makes combination therapy directed to distinct parasite targets a reasonable alternative. Musonda et al. developed hybrid molecules combining two structurally distinct moieties of chloroquine and astemizole [15]. *In vitro* efficacy of hybrids showed to be superior to chloroquine (IC$_{50:}$ 0.230 µM) against a chloroquine-resistant *P. falciparum* strain yielding IC$_{50}$ values between 0.037 and 0.064 µM and high selectivity index (SI) ranging from 142 to 298. Furthermore, two of these hybrid analogues possessed significant *in vivo* activity in mice infected with chloroquine-resistant *P. berghei* (a rodent *Plasmodium*). A 4-day treatment using doses of 20-50 mg/kg ip did not prolong the mean survival time compared with chloroquine (8-13 days with analogues vs. 17 days with chloroquine), but markedly reduced parasitemia [15]. Consequently, the authors specified that further efforts were necessary to improve the drug-like properties of this class of compounds.

Identifying drugs capable of targeting the liver stages and principally the *P. vivax* hypnozoite responsible for persistent infection is crucial for successful malaria control programs. Interestingly, the activity of astemizole also comprised the parasite's liver stage, as shown in a screening assay using HepG2 liver cells infected with luciferase-transfected *P. berghei* sporozoites harvested from mosquitoes [17] (Supp. Table 1). The HepG2 cytotoxicity (IC50: 12 µM) and antiplasmodial activity measured by luminometry (IC50: 0.114 µM) produced an excellent SI (105.3) suggesting astemizole potential for *P. vivax* therapy in combination with other antimalarial drugs. However, this should be interpreted with caution because *P. berghei* does not produce hypnozoites. Consequently, the drug should be further evaluated against these forms, which can be produced in primary cultures of hepatocytes infected with *P. vivax* [18, 19] or *P. cynomolgus* (a *Plasmodium* species from non-human primates) [20]. Alternatively, i*n vivo* evaluations could be performed in liver humanized FRG KO huHep mice, which have shown to be useful for *P. vivax* studies that include the hypnozoites forms [21].

Utilizing the inhibition of lactate dehydrogenase (LDH) obtained from *P. falciparum* grown in culture medium as a surrogate of anti-plasmodial activity, Roman et al. showed that astemizole and some of its derivatives had $IC_{50}$ values within the nanomolar range ($IC_{50}$<0.4 µM) [16] (Supp. Table 1). Two of these analogues exhibited low cytotoxicity for Chinese hamster ovary cells (CHO) resulting in high SI (>120). The authors indicated that the benzimidazole core of astemizole is likely an essential structural requirement linked to antiplasmodial activity.

*In vivo* experiments utilizing astemizole (30 mg $m^2$/day) or desmethylastemizole (15 mg $m^2$/day) in mice infected with a chloroquine sensitive strain of *Plasmodium vinckei* (a rodent parasite) yielded 80% reduction in parasitemia at day 5 post-treatment compared to untreated mice [22]. The same *in vivo* experimental protocol performed with a chloroquine-resistant strain of *P. yoelii* (another rodent parasite), reduced parasitemias by 44% with astemizole and 40% with desmethylastemizole. Astemizole was only parasitostatic at 15 mg $m^2$/day, as indicated by parasite recrudescence after 4 days of suspending treatment, but curative results were achieved after a 4-day treatment using a significantly higher oral dose (300 mg $m^2$/day) [22].

Tian et al. synthetized astemizole analogues with the two-pronged objective of identifying molecules with antiplasmodial activity that also possessed lower hERG inhibitory properties associated with rare but serious cardiac side effects linked to high doses [23, 24]. Some analogues displayed $IC_{50}$ values of 30 nM against *P. falciparum* and showed reduced hERG inhibition compared to astemizole. However, the best analogue achieving significant antiplasmodial activity and low hERG affinity was markedly lipophilic (clogP = 7.7), a property associated with low solubility, frequent biotransformation, and potential drug-drug interactions [25]. Therefore, additional medicinal chemistry work will be essential to improve the aqueous solubility while reducing lipophilicity of newer astemizole analogues.

Recent studies involving synthesis of other astemizole analogues found that many of them (10/17) also showed submicromolar activity and high SI (>100) against chloroquine sensitive and multidrug resistant *P. falciparum* [22]. Interestingly, evaluations of the antiplasmodial activity of these analogues included gametocytes (IC50: 1−5 µM), which are essential for parasite multiplication within mosquito vectors (Supp. Table 1). Furthermore, one of the analogues was 3-fold more active than astemizole against the liver stage of *P. berghei*. As expected, mechanistic studies using these astemizole derivatives found a positive correlation between inhibition of parasite detoxification of heme (through hemozoin formation) and antiplasmodial IC$_{50s}$. Notably, like other astemizole studies no preclinical evaluations using experimental animals have been reported.

Clemastine, a first-generation antihistamine originally identified by Derbyshire et al. as active against the liver stage of *P. berghei*, was further studied *in vitro* using parallel chemoproteomics to unravel its mechanism of action (described below; Lu et al., 2020)[17]. The drug was active at low concentrations against the blood forms (trophozoite: 2.3 ± 0.33 µM) and liver-stage (merozoite: 9.3 ± 1.1 µM) at 24 hours post-drug exposure.

Clemastine binds and destabilizes *P. falciparum* chaperonin containing TCP-1 (TRiC/CCT), which is required for *de novo* cytoskeletal protein folding of the parasite. In addition, *P. berghei* exposure to clemastine resulted in alteration of mitotic spindles during the asexual reproduction, leading to abnormal tubulin morphology specific to *Plasmodium* but not to human cells. This specific mechanism of action against *Plasmodium* should aid the screening of new compounds that impair protein folding to treat malaria [26]. Like quinolones, the mechanism of action of astemizole and desmethylastemizole relates to the impairment of heme crystallization (hemozoin) and progressive accumulation of toxic heme within the *P. falciparum* food vacuole in both chloroquine-sensitive and multidrug resistant strains [27].

Overall, several astemizole analogues have shown capacity to inhibit all stages of *Plasmodium* spp. However, it is disappointing that no comprehensive *in vivo* trials have provided definitive answers of its therapeutic potential. Therefore, it is imperative that additional medicinal chemistry and *in vivo* trials are undertaken to improve compound drugability and confirm their applicability in antimalarial therapy. Clemastine demonstrated excellent activity *in vitro* and its mechanism of action was well defined, nonetheless like in the case of astemizole and it analogues, preclinical studies still need to be performed for assessing its purported antimalarial efficacy. The capacity of these two antihistamines to inhibit the dormant liver stage of *P. vivax* is another pending issue that could be evaluated in liver humanized FRG KO huHep mice [21].

**3.2. Leishmania spp.**

*Epidemiological importance*. Leishmaniasis is produced by different species of *Leishmania* and represents one of the most important neglected tropical diseases after malaria. It is transmitted by the bite of infected sand flies and depending on the *Leishmania* species it can produce cutaneous, muco-cutaneous, or visceral disease. Visceral leishmaniasis (VL) produced by *L. donovani* or *L. infantum* could be life threatening if left untreated, but is less common than cutaneous leishmaniasis, and produces 50,000-90,000 annual cases [28]. Approximately 95% of cutaneous leishmaniasis (CL) cases occur in the Americas, the Mediterranean basin, the Middle East and Central Asia. There is no accurate data on the number of infected people because underreporting is a common problem; however, WHO estimates that between 600,000 and 1 million new annual cases of cutaneous leishmaniasis occur worldwide [28].

*Current treatment and mechanism of action*. Several drugs have shown (variable) efficacy for treating leishmaniasis, but the more widely used drugs are pentavalent antimonials for CL, liposomal amphotericin B for VL and miltefosine against both diseases [29]. The increasing parasite drug resistance and toxicity of current therapies are the major justifications for identifying new antileishmanial compounds or combination therapies [30]. The mechanism of action of pentavalent antimony has not been fully elucidated, but it has been shown that inhibition of *Leishmania* DNA topoisomerase I and perturbation of energy metabolism are in part responsible for parasite killing [31-34]. Miltefosine mechanism of action is principally based on the oxidase-stress that induces *Leishmania* apoptosis [35]. Liposomal amphotericin B seems to bind to *Leishmania* ergosterol precursors, e.g. lanosterol, causing disruption of parasite membrane and loss of the permeability barrier to small metabolites [34, 36].

*Antihistamines and mechanism of action.* Repurposing of FDA-approved drugs has become a predominant strategy for the identification of candidate molecules against *Leishmania* [37]. Using this approach, Peniche et al. screened eleven H1R antagonists using an *ex vivo* explant culture of lymph nodes from mice infected with luciferase-transfected *Leishmania major* (*Luc-L. major*) [38]. The work identified azelastine (AZ) and fexofenadine (FX) as potential antileishmanial molecules due to their significantly low $EC_{50}$ (of 0.05 µM and 1.50 µM, respectively) against intracellular amastigotes (Supp. Table 1). Furthermore, a cell-based assay using the HepG2 cell line (human hepatocellular carcinoma) indicated that both compounds had low cytotoxicity, yielding high *in vitro* therapeutic indexes (= selectivity index of 942 for AZ and 541 for FX) [38]. AZ markedly reduced the parasite burden in mouse peritoneal macrophages infected with *L. major* suggesting good cell penetration. The study also showed that AZ promoted macrophage immune responses that enhanced parasite clearance.

In vivo studies in Balb/c mice infected with *Luc-L. major* confirmed the *ex vivo* observations described above. Infected mice treated orally (80 mg/kg bid) for 10 days with FX or using three intralesional injections of AZ had a significant reduction of lesion size (FX = 69%; AZ = 52%), as well as significant parasite suppression in the lesion (FX = 82%; AZ = 87%) and draining lymph nodes (FX = 81%; AZ = 36%) [38]. So far, no mechanism of action of both antihistamines against *Leishmania* has been investigated.

Pinto et al. evaluated several antihistamine drugs against *L. infantum* promastigotes, but most of them displayed antileishmanial activity at *in vitro* concentrations of 13–84 µM that are usually considered high for hit selection [39]. In fact, only cinnarizine eliminated 100% amastigotes from peritoneal macrophages after 120 h of exposure, yielding an $IC_{50}$ of 21.53 µM (Supp. Table 1). No selectivity index was provided. The liposomal formulation administered i.p. at 3 mg/kg for eight days reduced *L. infantum* load to 54% in the liver of infected hamsters as determined by real time PCR but had no effect on spleen parasites. The authors hypothesized that the presence of a piperazine ring, like in other antileishmanial drugs (i.e. ketoconazole), could be responsible for cinnarizine activity. Therefore, they considered that medicinal chemistry approaches could use cinnarizine's drug scaffold to produce novel analogues for further antileishmanial screening [39].

Cinnarizine acts as a calcium channel blocker and may be linked to its mode of action against *Leishmania* [40]. Additionally, De Melo Mendes et al. found that cinnarizine was active against *L. infantum* amastigotes at an $IC_{50}$ of 19.9 µM as determined in spleen macrophages from hamsters [41]. Measurement of mitochondrial membrane potential of promastigotes using the fluorescent probe Rh 123 and transmission electron microscopy established that *L. infantum* killing was due to intracellular disorganization, nuclear membrane detachment, mitochondrial damage, and dysfunction of the respiratory chain [41] (Supp. Table 1).

More recently, clemastine was identified through *in vitro* screening as a drug candidate against multiple *Leishmania* species [42]. Clemastine is an ethanolamine-derivative, first generation antihistamine that binds to the H1R and has anticholinergic and sedative effects [43]. Clemastine fumarate showed significant levels of activity against promastigotes as determined by its markedly low $EC_{50}$s against *L. major* (0.179 µM), *L. amazonensis* (0.028 µM), *L. donovani* (0.019 µM) and *L. infantum* (0.708 µM). Furthermore, its antileishmanial activity was confirmed in *L. amazonensis*-infected macrophages in which amastigote suppression was achieved at a submicromolar concentration ($EC_{50}$ = 0.462 µM) and displaying good selectivity index (SI=39.49) based on its cytotoxicity data ($CC_{50}$= 18.25 µM) [42] (Supp. Table 1). Clemastine showed potent inhibition of inositol phosphorilceramide synthase, which is essential for the synthesis of *Leishmania* sphingolipids. In addition to this metabolic perturbation, it was proposed that interaction with the macrophage P2X7 receptor (ligand-gated ion channel) could be in part responsible for the antileishmanial activity [42].

The high parasite sensitivity to clemastine fumarate found *in vitro* was subsequently confirmed in Balb/c mice. Intraperitoneal administration of clemastine to Balb/c mice infected with *L. amazonensis* at a dose of 11.65 mg/kg/ip, twice a week for 28 days resulted in significant decrease in lesion size and parasite load at 41 days pi and proved to be superior to the intralesional or oral routes [42]. No complete parasite clearance was achieved, but this is not unexpected after only seven days of completing treatment. Overall, the data suggested that higher doses or combination therapy with antileishmanial drugs will be necessary.

Taken together, the studies indicated that several antihistamines have *in vitro* an *in vivo* activity against different *Leishmania* species. Some of them altered parasite metabolism and may be promoting favorable host immune responses. The utilization of parasite targets different to those currently used by antileishmanial drugs place antihistamines as potential therapeutic tools for combination therapy to reduce parasite drug resistance and lower toxic regimens (i.e. pentavalent antimony and miltefosine). Enhancement of antihistamine analogue synthesis, better identification of parasite target(s) and expansion of preclinical *in vivo* studies are still necessary before human trials could be recommended.

### 3.3 Trypanosoma cruzi

*Epidemiological importance*. The etiological agent of American trypanosomiasis (Chagas Disease) is *Trypanosoma cruzi*, a protozoan parasite transmitted through the feces of infected triatomine bugs (kissing bugs) during blood feeding on humans. The parasite is widely distributed in the American continent from southwestern US to northern Argentina and Chile. Approximately 20 million people are at risk and around 6 million are currently infected. It is estimated that >10,000 deaths occur annually due to cardiomyopathies or megaviscera complications [44].

*Current treatment and mechanism of action*. It is based on two toxic drugs, nifurtimox and benznidazole, with common side effects such as nausea, vomiting, dermatitis, and peripheral and central nervous system disorders. In search for efficacious and less toxic drugs, triazole derivatives were considered good alternatives as suggested by preclinical studies. However, clinical trials using posaconazole alone or in combination with benznidazole showed no advantages over benznidazole monotherapy [44]. This was probably unexpected since the mode of action of both drugs is different. Benznidazole produces lethal double-stranded DNA breaks in *T. cruzi*, while posaconazole (in yeasts) interferes with the enzyme that synthesizes ergosterol, which is essential for cell membrane stability and transport of nutrients [45, 46]. The mechanism of action of nifurtimox is not fully established. The drug activates nitroreductase enzymes that produce reactive metabolites toxic for *Trypanosoma cruzi*. The inhibition of *T. cruzi* dehydrogenase activity is another probable mode of action of nifurtimox that needs further studies [47].

*Antihistamines and mechanism of action.* The high toxicity of existing drugs and lack of efficacious alternative therapies makes searching for new parasite targets through drug screening a sensible approach. Screening of FDA-approved molecules using murine 3T3 fibroblasts infected with *T. cruzi* identified clemastine as a potential drug against this parasite [48]. The drug showed an *in vitro* EC$_{50}$ in the 0.37-0.44 µM range, but without apparent *in vivo* activity. Nevertheless, *in vitro* determinations using isobologram analysis showed that the combination of clemastine with cloperastine (an antitussive drug with H1R antagonist activity) resulted in synergistic activity *in vitro* based on 93% parasite suppression compared with inhibitions by cloperastine (21%) or clemastine (26%) alone [48]. The combination of clemastine with posaconazole was not synergistic nor it cleared *T. cruzi* but caused significant decrease in parasitemia and prolonged survival of infected mice compared to posaconazole used as monotherapy (Supp. Table 1). Although the mechanism of action of clemastine against *T. cruzi* was not addressed in this study, it is reasonable to assume that like in other kinetoplastid organisms (e.g. *Leishmania*) perturbation of lipid metabolism could be involved in parasite suppression [49].

Surprisingly, no additional studies using clemastine are available since the original work was published. An important issue regarding Chagas disease treatment is the time at which therapy is initiated. Nifurtimox or benznidazole treatment of acute infections in adults or children during the first year of life is associated with high rates of therapeutic success compared with chronically infected individuals [50]. The applicability of small animal models to assess the efficacy of new drugs is unclear since results vary depending on the mouse/*T. cruzi* strain combination. Nevertheless, it has been claimed that mice could be useful to recreate treatment response during the acute and chronic phase of Chagas disease [51]. Additionally, we believe that dogs could be an interesting option to evaluate prevention of cardiomyopathies due to their propensity to develop heart pathologies upon natural *T. cruzi* infection [52].

### 3.4 Schistosoma spp.

*Epidemiological importance.* Schistosomiasis is a tropical parasitic disease caused by trematode worms of the genus *Schistosoma* that affects approximately 200 million people and accounts for an estimated 280,000 annual deaths in sub-Saharan Africa [53]. The life cycle encompass fresh water snails (intermediate host) and vertebrate animals (definitive hosts) [53]. Humans become infected when working or bathing in water where infective *Schistosoma* larvae (cercariae) are released from infected snails. Male and female schistosomes live in copula in the venous plexus (*S. haematobium*) or the mesenteric veins (*S. mansoni, S. japonicum*) of the host and the eggs released by -the female become trapped in the tissues. This elicits an immunopathologic reaction leading to intestinal disease, liver fibrosis or obstructive disease in the urinary tract, depending on the infecting *Schistosoma* species [53].

*Current treatment and mechanism of action*: Praziquantel is the drug of choice for treating schistosomiasis but fear of increasing parasite drug resistance has been justified [54]. In fact, selection of praziquantel-tolerant schistosome strains in lab animals was achieved without difficulty [55]. Drug tolerant strains were found in Egypt and treatment failure was reported in Senegal, although further clarification of other potential intervening factors need to be investigated [53]. Synthesis of 19 praziquantel analogues resulted in molecules with moderate activity against schistosomes, yet none of them as active as the parent drug [56]. Therefore, there is a justifiable urgency to identify new drugs that could complement the existing treatment, which may be at risk of becoming less efficacious.

Several mechanisms of action have been proposed but not confirmed, ranging from $Ca^{2+}$ influx that disrupts the parasite membrane, alterations of the lipid bilayer that surrounds schistosomes, to inhibition of sphingomyelinase activity resulting in impaired reproduction and egg release [54]. Consequently, the lack of well-defined parasite targets makes the screening for novel drugs or design of praziquantel analogues more difficult.

*Antihistamine and mechanism of action.* Recent *ex vivo* assays demonstrated that promethazine was able to kill adult *S. mansoni* with 50% lethal concentration (LC50) of 5.84 μM after72 h of drug exposure. The drug interfered with parasite motility and viability, and produced significant damage to the schistosome tegument [57]. Promethazine treatment of mice using an oral dose of 100 mg/kg x 5 days at different post-infection intervals (prepatent and patent infections) demonstrated that promethazine was active against juvenile and adult worms (Supp. Table 1). The anti-schistosomal activity was confirmed by the significant reduction in worm loads, egg production and diminution of hepatomegaly and splenomegaly of treated mice. The highest promethazine efficacy in worm burden reduction was achieved in patent infections (>90%). An important finding that could limit parasite transmission was the marked egg reduction in feces (96.1%; P< 0.0001) of promethazine therapy compared with untreated control mice [57].

Kimber et al. cloned a functional G protein-coupled acetylcholine receptor called AsGAR-1 from the pig gastrointestinal nematode *Ascaris suum* and found that promethazine reduced AsGAR-1 activation to 10 ± 3.5% of control [58]. Subsequently, MacDonald et al. described a functional G protein-coupled acetylcholine receptor in *S. mansoni* (SmGAR), which is constitutively active and could be further activated by acetylcholine [59]. On the other hand, suppression of SmGAR using RNAi in a schistosomulae phenotypic assay showed a marked diminution in larval motility indicating that *Schistosoma* SmGAR is a promising drug target. In agreement with the findings on ascaris AsGAR, promethazine also acted as an inverse agonist of SmGAR leading to significant decrease in the receptor's basal activity.

The *in vitro* and *in vivo* observations of different studies underscore the relevance of the acetylcholine receptor GAR as target for *Ascaris* sp. and *Schistosoma* sp. and point to the possible therapeutic importance of promethazine, thereby paving the road for additional preclinical studies with this first-generation antihistamine.

### 4. Activity against fungi and bacteria

#### 4.1. *Cryptococcus spp.*

*Epidemiological importance. C. neoformans* is the most common cause of meningoencephalitis in adults living with HIV, principally in sub-Saharan Africa, and causes 15% of AIDS-related deaths globally [60]. This dimorphic fungus is widely spread in the environment and spore inhalation is typically asymptomatic, as *Cryptococcus* is either eliminated by the immune response or becomes dormant in immunocompetent individuals [60]. However, it is a leading cause of disease and mortality in immunocompromised individuals resulting in >220,000 infections and approximately 180,000 deaths annually [61].

*Current treatment and mechanism of action. Cryptococcus* spp. are susceptible to three classes of antifungal agents, including polyenes (amphotericin B), flucytosine (5-fluorocytosine), and azoles. Although the mechanism of amphotericin B remains unclear, it was suggested that it perturbs ergosterol metabolism since mutations in an enzyme implicated in its biosynthesis (sterol Δ8-7 isomerase) correlated with decreased ergosterol content in a resistant *C. neoformans* isolate [61]. Fluocytosine is converted to 5-fluorouracil by the enzyme cytosine deaminase within the fungus disrupting RNA synthesis and subsequent inhibition of DNA synthesis through its conversion to fluoro-deoxyuridylic acid [62]. Azoles directly impair ergosterol biosynthesis of *Cryptococcus* sp. by inhibiting the activity of lanosterol 14α-demethylase [63]. Notably, approximately one third of patients with cryptococcal meningitis subjected to standard therapy fail to resolve the infection stressing the need for new, more active drugs [64].

*Antihistamine and mechanism of action.* Utilization of disk diffusion assays in agar plates infected with *C. neoformans* showed that fluconazole, which is considered to be fungistatic, displayed fungicidal activity against *C. neoformans* in combination with the antihistamine astemizole. Agar halos clear of *C. neoformans* were found around the 6mm filter papers loaded with 32µg of fluconazole alone while the same results were obtained with 8 µg of fluconazole in combination with 32 µg of astemizole [65]. A similar result was achieved with one of three astemizole analogues included in the assays (Supp. table 2). Furthermore, the combination with the most active analogue (A2) and fluconazole against *C. gattii* lowered the minimum inhibitory concentration (MIC) of fluconazole almost 10-fold. The fractional inhibitory concentration index indicated that both astemizole and A2 had synergistic *in vitro* fungicidal activity[65].

An established *Galleria mellonela* (wax moth) system was used to determine the *in vivo* efficacy of these antifungal agents [65, 66]. Larvae previously challenged with *C. neoformans* showed improved survival upon combined treatment with fluconazole and astemizole or fluconazole and A2 (doses not described) compared with untreated larvae. By the study endpoint (6.5 days) the Kaplan-Meier analysis showed that approximately 19% (P=0.019) of caterpillars treated with fluconazole+astemizole or 28% (P=0.002) treated with fluconazole+A2 survived compared to none of the controls. Conversely, no *in vitro* synergy or improved survival was observed in caterpillars challenged with a fluconazole-resistant strain of *C. neoformans* (89-6105797)[65]. It will be necessary to identify the mechanism of action of these combinations for the development of novel astemizole analogues with improved combination potency against drug-resistant *Cryptococcus* species. *G. mellonella* is mostly used to evaluate the innate immune response to pathogens, however the humoral response is not representative of human immune response [66]. Therefore, confirmation of fluconazole+astemizole activity in an established mouse model of meningoencephalitis will be essential to obtain relevant preclinical data [67].

### 4.2. Candida spp.

*Epidemiological importance.* Candidiasis is a common fungal disease that has been considered of minor medical importance affecting mostly women (vaginosis) or babies (thrust). However, epidemiological data showed that infants with low birth weight (<1000 g) are at high risk of invasive candidiasis and death [68]. Neonatal candidiasis has been related to 20% mortality, and frequent damage to heart, genitourinary tract and central nervous system [69]. Furthermore, invasive candidiasis due to *C. albicans*, *C. glabrata*, *C auris* or *C. parapsilosis* is a major problem in ICU, and is more commonly associated with surgery, prolonged candidaemia and immunosuppression [70].

*Current treatment and mechanism of action.* Treatment of invasive candidiasis encompass distinct drug classes: polyenes, azoles, echinocandins and also flucytosine [71]. The polyene amphotericin B binds to ergosterol of the fungal cytoplasmic membrane creating aqueous pores that lead to ion leakage and cell lysis. Azoles are fungistatic, inhibiting Candida's lanosterol 14α demethylase thereby depleting ergosterol, which is the principal plasma membrane sterol in fungi. The echinocandins inhibit the transmembrane glucan synthase complex (Fks1) disrupting the synthesis of 1,3-β-D-glucan damaging the cell-wall carbohydrate leading to cell rupture and death [71].

*Antihistamines and mechanism of action.* Dennis et al. utilized checkerboard assays to evaluate the anti-*Candida* activity of the 2$^{nd}$-generation antihistamines terfenadine (TERF) and ebastine (EBA) in combination with seven azoles and against 13 *Candida* strains (Supp. table 2) [72]. The inhibition of fungal growth was evaluated in infected RPMI 1640 medium using time-kill assays, while additional studies encompassed the cytotoxicity to mammalian cells and the capacity of drug combinations to disrupt biofilm (sessile) formation in 96-well plates. They found synergistic activity in several combinations of azoles with TERF (n=41) or EBA (n=14). Furthermore, TERF and azole combinations displayed synergy against *C. albicans* including strains resistant to most triazoles. Combinations at variable posaconazole (POS) concentrations were done at a fix EBA dose of 3.1 µg/mL since this drug concentration had no cytotoxicity for all the cell lines evaluated. While no biofilm activity against sessile *C. albicans* was observed in the posaconazole (POS)-EBA combination, it proved to be synergistic against *C. glabrata* biofilm decreasing the sessile MIC (SMIC) approximately 500-fold for POS and 4-fold for EBA. *In vitro* cytotoxicity assays showed that cell survival rates of POS plus EBA (at 3.1 µg/mL) were not significantly different from POS alone in all the four cell lines evaluated. Based on structural similarity, it was hypothesized that TERF and EBA should have the same mechanism of action, however the fungal targets still must be identified [72]. Despite neutropenic immunosuppressed mice have been used to evaluate drugs (e.g. rezafungin) against disseminated candidiasis or *Candida* spp. virulence, to date no preclinical trials with these antihistamines have been published [73, 74].

**4.3 Staphylococcus aureus**

*Epidemiological importance. S. aureus* is a frequent finding in the skin and nose of healthy individuals (carriers), but is also the most important cause of bacteremia in the United States that could lead to significant morbidity and mortality [75, 76]. Furthermore, invasive methicillin-resistant *S. aureus* (MRSA) infections are associated with infective endocarditis, osteomyelitis, and sepsis [77]. Treating MRSA bacteremia could be difficult and requires selection of antibiotics based on susceptibility patterns.

*Current treatment and mechanism of action.* Vancomycin, a glycopeptide antibiotic against gram-positive bacteria has been the main drug used to treat *S. aureus*. Vancomycin binds to bacterial D-Ala-D-Ala enzyme preventing cell wall synthesis of the long polymers of *N*-acetylmuramic acid (NAM) and *N*-acetylglucosamine (NAG), which constitutes the principal structure of the bacterial cell wall. However, bacteria can become vancomycin resistant by substituting the terminal D-alanine of the peptidoglycan precursor with D-lactate impairing antibiotic binding to the cell wall [78]. Although vancomycin analogues and new drugs are used in combination with vancomycin, there is increasing inefficacy of available antibiotics underscoring the need to find new drugs.

*Antihistamines and mechanism of action.* A phenotypic high-throughput screening using an off-patent FDA- approved chemical library identified the second-generation antihistamine terfenadine (TRF) to be active against methicillin-resistant *Staphylococcus aureus* (MIC = 16 µg/mL) and other gram-positive bacteria [79]. In that work, 84 TRF-based analogues were synthesized to optimize the activity against *S. aureus*. While two of the analogues improved their antibacterial activity compared to TRF (MIC= 1-4 µg/mL) there was no improvement regarding hERG (inhibition of the human ether-a-go-go related gene [hERG] potassium´ channel) liability for cardiac side effects, with inhibitory activity at sub-micromolar concentrations. *In vitro* experiments in this study indicated that the most probable antibacterial mechanism of action of TRF and its analogues corresponded to the inhibition of bacterial type II topoisomerases (DNA gyrase and topoisomerase IV). Additionally, the authors hypothesized that the high lipophilicity (logP) of TRF analogues could favor interactions with multiple targets, since receptor promiscuity is correlated with lipophilicity [79, 80]. At present, no *in vivo* studies with TRF or any of the analogues have been published.

Additionally, Cutrona et al. found that loratadine (LRT) potentiates both β-lactam antibiotics and vancomycin against methicillin-resistant (MRSA) and vancomycin-resistant *S. aureus* (VRSA) [81]. Treatment with LRT at 50 µM was effective against the clinically relevant MRSA strains (USA100 and USA300), reducing the MIC of oxacillin 1024 and 64 folds, respectively. LRT exhibited marked biofilm inhibitory activity against all strains, with minimum biofilm inhibitory concentration (MBIC50) values ranging from 11.49 to 59.90 µM in both *S. aureus* and *S. epidermidis* strains (Supp. Table 2). This antihistamine was not able to potentiate oxacillin in MRSE strains of *S. epidermidis*, but displayed 8-fold reduction in the oxacillin MIC against VRSA strains indicating that it may be interfering with an antibiotic resistance pathway specific to *S. aureus*. LRT targeted the regulatory serine−threonine kinases Stk1 (PknB) in *S. aureus* and Stk in *S. epidermidis* altering the transcription of genes related to biofilm formation and antibiotic resistance in both bacterial species [81]. From the safety standpoint it is worth noting that LRT has an excellent profile in children and adults even at concentrations 30-fold higher than the recommended dosage [82, 83]. No preclinical trials using antibiotic-antihistamine combinations were described, although other authors have used rat and rabbit models to evaluate antimicrobials in *S. aureus* infections [84, 85].

## 5. Activity against viruses

Viral diseases are responsible for devastating pandemics (e.g. HIV, SARS-coV-2, Influenza) and epidemic or chronic diseases (e.g. Ebola, Hepatitis C, RSV, Herpes), all of them difficult to control or treat. Antiviral drugs curb virus infections using different mechanisms. It encompass inhibition of virus attachment, entry, uncoating, or blockade of different viral enzymes such as polymerases, proteases, nucleosides, nucleotide reverse transcriptases and integrases [86].

The search for new drugs that are efficacious and less expensive than molecules identified through traditional discovery approaches involve the screening of FDA-approved drugs that could interfere with any of these viral targets. Among other molecules, several antihistamines have demonstrated the capacity to impair viral infection and replication.

### 5.1. SARS-coV-2

*Epidemiological importance.* According to NIH, as of February, 2022, the SARS-CoV-2 pandemic accounted for more than 426 million cases and more than 5.8 million deaths, globally [87]. While vaccines have shown to be effective in preventing serious disease, regardless of virus variants, unvaccinated and individuals with co-morbidities (obesity, diabetes, elderly) are at higher risk may still develop severe pathology and antiviral therapy is indicated [88].

*Current treatment and mechanism of action.* The current small molecules authorized or approved for SARS-coV-2 treatment are the orally administered viral protease inhibitor nirmatrelvir (+ritonavir) and the viral RNA-polymerase lethal mutagenic drug molnupiravir. Remdesivir, which interferes with RNA transcription prematurely, is delivered intravenously and therefore can be used only in hospital settings [87] The novel oral antivirals molnupiravir, fluvoxamine and Paxlovid showed to be effective in decreasing mortality and hospitalization in COVID-19 patients and exhibited good overall safety [89]. However, there is still much interest is the identification of drugs that could increase the therapeutic options against SARS-coV-2.

*Antihistamines.* Based on the notion that the virus (Delta variant) enters the host cells using the angiotensin-converting enzyme 2 (ACE2) receptor, Ge et al. used cell membrane chromatography (CMC) as an *in vitro* method to screen histamine H1 receptor antagonists (H1Ra) that could bind to ACE2 receptor [90]. Doxepin (DXP), an H1Ra with antidepressant and anxiolytic properties, showed good affinity for the ACE2 receptor and also inhibited in a dose-dependent manner the entry of a SARS-coV-2 pseudovirus into cells (HEK293T) that overexpress ACE2. Utilization of a non-cytotoxic concentration of doxepin (20μM) resulted in approximately 74% inhibition of virus entrance into HEK293T cells [90]. Using the same experimental approach it was determined that the H1Ra antihistamines azelastine, loratadine and desloratadine had affinity for the ACE2 receptor and inhibited pseudovirus entry into HEK293T cells, as determined by luminometry [90, 91]. At 20μM concentration, azelastine had good antiviral activity (EC50 = 3.8 μM) and acceptable cytotoxicity (CC50= 84.1 μM) yielding a selectivity index of 21.95 [90]. Evaluations using plasmon resonance showed that desloratadine had higher affinity for the ACE2 receptor than loratadine and resulted in approximately 70% inhibition of virus entrance into HEK293T cells [91, 92]. Not surprisingly, other screening methods against SARS-coV-2 have identified additional antihistamines of which clemastine demonstrated to be the most active drug (Yang et al., 2021). Clemastine exhibited potent EC90 (2.01 ± 0.94 μM) and EC50 (0.95 ± 0.83 μM) values and a selectivity index greater than 20 when evaluated at a 20 μM concentration. The study showed that clemastine inhibited the entry of SARS-CoV- (and MERS-CoV-S) pseudotype virus in Huh-7 cells [93].

A retrospective study that used logistic regression analysis from data of 219,000 persons who regularly used antihistamine drugs vs. individuals that did not, found an association (odds ratio) between prior prescription of antihistamines and negative SARS-CoV-2 test results as primary outcome adjusted for sex and age. Results showed that prior utilization of the antihistamines loratadine, diphenhydramine, cetirizine, hydroxyzine, and azelastine was associated with a significantly decreased incidence of SARS-CoV-2 test positivity in individuals ≥61 years old. For example, the strongest association between antihistamine use and covid-19 test negativity was observed with azelastine (n=135) with an OR=2.43 ±1.47-4.02 (p=0.001) [94].

Additional *in vitro* studies using VERO cells infected with a SARS-CoV-2 isolate showed that hydroxyzine, diphenhydramine and azelastine exhibited direct antiviral effects. However, except for azelastine the $EC_{50}$ values were above the plasma concentrations of clinically recommended doses, and therefore no further development or clinical applicability are feasible [94]. As with many of the antihistamine studies, no preclinical evaluations were published despite the availability of different mouse models and hamsters that have shown applicability for anti-SARS-coV-2 drug studies [95].

## 5.2. Ebola virus (EBV)

*Epidemiological importance.* Since 1976, there have been 34 recorded outbreaks of Ebola virus (EBOV) totaling >34,000 cases and 14,823 deaths, encompassing 11 countries in Sub-Saharan Africa. There is a wide range in fatality rate (26 to 88%) but the overall rate (95%CI) is 66% (62 to 71%) [96]. Marburg virus (MARV) is similar to EBOV and is responsible for outbreaks in Sub-Saharan Africa with fatality rates of 24 to 88% [97]. EBOV and MARV are single stranded RNA filoviruses that are classified as category A agents (NIAID) because they could be used as bioweapons. Consequently, the development or identification of highly efficacious drugs to treat these viruses is crucial.

*Current treatment and mechanism of action.* To date, therapy relies on monoclonal antibodies (mAb114 and REGN-EB) and the antiviral drug remdesivir that targets viral RNA-dependent RNA polymerase. However, further treatment optimization including development of newer small molecules is recommended due to the still unacceptable post-treatment mortality rate (64-70%) [98].

*Antihistamine drugs.* Screening of H1R antagonists for anti-filovirus activity identified diphenhydramine and chlorcyclizine as potential drugs for EBOV and MARV [99]. The initial screening was done at biosafety level 2 utilizing pseudoviruses constituted by a HIV replication-deficient core and glycoproteins of EBOV or MARV. Subsequent studies with live EBOV (Kikwit) were performed at biosafety level 4 facilities using human foreskin fibroblasts (HFF-1) and HeLa cells. The $IC_{50}$ of diphenhydramine was 5.7µM in HeLa cells and 2.2 µM in HFF-1 cells, while chlorcyclizine values were $IC_{50}$= 5.7 µM in Hela cells and 3.2 µM in HFF-1 cells. Utilization of the Luminescent Cell Viability Assay (Promega) in the human lung epithelial cell line A549 showed that both drugs had selectivity indexes >10. Chlorphenoxamine, a diphenhydramine analogue, showed an excellent selectivity index (>50) against EBOV (Supp. Table 3). On the other hand, the 2nd and 3rd generation antihistamines cetirizine and fexofenadine, respectively, had no anti-filovirus properties [99].

Infection of A549 cells with pseudo viruses and time-of-drug-addition experiments adding histamine or acetylcholine to saturate their receptors, suggested that the anti-filovirus mechanism of action was not related to interactions with these receptors. Instead, it pointed to the blockade of viral entry in the endosome preceding to fusion. Nevertheless, adhesion of these antihistamines to the viral glycoprotein as another antiviral mechanism could not be rule out [99]. The authors indicated that for potential applicability the development of more potent analogues will be necessary since the observed $IC_{50s}$ were much higher than the FDA-approved doses for current allergy therapy. Small animal models of filovirus do not accurately reproduce human disease, however rodents could still be considered for antihistamine efficacy studies using mouse- or hamster-adapted EBOV viral strains, as well as immunocompromised (SCID) or genetically modified mice [100].

### 5.3 Hepatitis C virus (HCV)

*Epidemiological importance.* Recent global estimates indicate that around 1.0% of the world population has viremic hepatitis C, which corresponds to 71 million active cases [101]. Chronic infection with the hepatitis C virus (HCV) could lead to renal impairment, cirrhosis, and hepatocellular cancer [102].

*Current treatment and mechanism of action.* A large proportion of currently used drugs are nucleotide/nucleoside inhibitors targeting NS5B, a viral polymerase involved in posttranslational processing vital for HCV replication [103]. These direct-acting-antivirals are orally active and can reach cure rates of approximately 95%. However, it would be important to identify new drugs that could help abbreviating treatment lengths to optimize compliance, reduce adverse effects, and decrease patient and payers' costs, principally in resource-limited countries [104, 105].

*Antihistamine drug.* Cell-based, high-throughput screening identified the 1[rst]-generation antihistamine chlorcyclizine HCl (CCZ) as a potent inhibitor of HCV [106]. The (S)-CCZ enantiomer had lower antihistamine activity than the racemic drug and consequently was chosen for further anti-HCV *in vitro* studies. The *in vitro* antiviral effect of (S)CCZ had negligible cytotoxicity and was highly synergistic in combination with anti-HCV drugs encompassing ribavirin, interferon-α, telaprevir, boceprevir, sofosbuvir, and daclatasvir as well as the immunosuppressant cyclosporin A [106].

In a luminometric HCV-Luc infection assay using human Huh7.5.1 hepatocytes, (S)-CCZ inhibited intracellular viral RNA at a low concentration (EC$_{50}$ 0.024 µM) and in a dose-dependent manner [106]. Mechanistic studies revealed that CCZ targets the fusion peptide of HCV E1 and restricts the fusion process, probably cross-linking to the E1 sequence next to the presumed fusion peptide, as determined by mass spectrometry analysis. Additionally, a binding model via docking simulations suggested that CCZ attaches to a hydrophobic pocket of HCV E1 leading to wide interactions with the fusion peptide [107].

The inhibitory kinetics of CCZ analogues on HCV infection was addressed in an *in vitro* time-of-addition experiment suggesting that the drug inhibited the early stage of viral entry into host cells. Additionally, the mouse pharmacokinetic model indicated that CCZ tends to accumulate in the liver, which is particularly important for treating this disease. Furthermore, treatment of chimeric mice engrafted with primary human hepatocytes showed no evidence of developing CCZ resistance. It was concluded that CCZ constitutes an excellent drug for repurposing, probably in combination with existing antiviral agents [106].

Additional studies by Rolt et al. focusing on the development of CCZ analogues, found that one of the compounds had potent *in vitro* anti-HCV activity (EC50= 2.3±0.9 nM) and excellent selectivity index (SI= 8609) (Supp. Table 3) [108]. Furthermore, as a "proof of concept", daily administration of 5 mg/kg IP for 4 weeks of this compound in severe immunodeficient mice (Alb-uPA/SCID/bg) infected with HCV showed a time-dependent viral load decrease of ≈1 log (=90%) in 3/4 mice. No evidence of toxicity was observed. In this study, experimental treatment of mice with the HCV protease inhibitor asunaprevir (not FDA approved) produced an initially strong antiviral activity followed by virus rebound, suggesting the development of drug-resistance. This rebound was not observed in 3 of the 4 mice treated with the CCZ analogue. Interestingly, pharmacokinetic data in mice demonstrated that the parent drug and its metabolite accumulated in the liver at levels much higher than plasma, which is a desired trait of anti-HCV compounds [108].

A small study involving CCZ treatment of 24 adult patients with a median HCV RNA load of 6.3 log IU/mL indicated that the drug used as monotherapy (2mg/kg x 28 days) produced no significant reduction in viremia (p= 0.69) [109]. However, 7/12 (58%) patients treated with CCZ+ ribavirin (RBV) had >3-fold decrease in HCV RNA. In a previous limited group of patients, RBV monotherapy was followed by virus load rebound while CCZ+RBV had similar efficacy in early viral suppression (58%) and sustained viral diminution suggestive of synergistic effect. There were only mild to moderate adverse effects not requiring treatment discontinuation. The fact that CCZ and ribavirin have different viral targets makes the combination of therapeutic significance. The authors suggested that larger CCZ doses could increase treatment efficacy, although toxicity could be a problem that will have to be assessed. First generation antihistamines are more proarrhythmic than second generation antihistamines but none of them are void of these serious and rare side effects [110]. Assessing toxicity of large doses requires preclinical studies in animal models such as dogs, monkeys, swine, rabbits, ferrets, and guinea pigs. Unfortunately, rats and mice are not useful due to their different mechanism of heart re-polarization compared to humans [111, 112].

### 5.4 Influenza virus

*Epidemiological importance.* Seasonal pandemic influenza affects individuals of all ages, but the elderly or immunocompromised persons and principally children are at higher risk of hospitalization and death [113]. A recent epidemiological review estimated that, globally, influenza virus was responsible for 10.1 million cases of acute lower respiratory infections (ALRI), 870,000 of ALRI hospitalizations, and 34,800 of ALRI deaths in children under 5 years. Approximately 82% of in-hospital deaths occurred in low-income and lower-middle income countries where vaccination rates are typically low [114].

*Current treatment and mechanism of action.* There are several drugs to curtail influenza infection interfering with different targets such as virus neuraminidase (e.g., oseltamivir, zanamivir) or virus polymerase (e.g., baloxavir marboxil) [115]. However, there is evidence of increased drug resistance because of their wide use (e.g. oseltamivir). Furthermore, these drugs are available mainly in developed countries but mostly inaccessible in developing countries, underscoring the need to identify new affordable and broadly obtainable drugs through screening of FDA-approved chemical libraries.

*Antihistamines.* Xu et al. screened an FDA-approved drug library composed of 1,280 compounds using a commercial cell assay (CCK-8) that measures the reduction of cytopathic effect (CPE) to identify anti-Influenza drug candidates [116]. Two 1rst generation antihistamines, carbinoxamine maleate (CAM) and S-(C)-chlorpheniramine maleate (SCM) demonstrated strong antiviral activity *in vitro* against an H7N9 strain with $IC_{50}$ of 3.56 and 11.84 µM, respectively (Supp. Table 3). Additional *in vitro* evaluations using CPE reduction assay indicated that CAM and SCM could inhibit infection by other influenza A H1N1 viruses, and one influenza B virus, displaying $IC_{50}$ and selectivity indexes ranging between 4.2-24.7 µM and 11.5-83.5, respectively. *In vivo* experiments using mice challenged intranasally with an H7N9 virus and treated IP with CAM (10 mg/kg per day) or SCM (10 mg/kg per day) for 5 days were fully protected [116]. *In vitro* time-of-addition studies indicated that both drugs blocked viral entry into cells but did not block virus attachment (HA activity) or virus release (NA activity), strongly suggesting they impair viral endocytosis (Xu et al. 2018). These encouraging results will have to be complemented by preclinical studies using oral or intranasal drug administration as opposed to invasive intraperitoneal delivery to confirm the potential applicability of these antihistamines for influenza treatment. Utilization of mice requires careful selection of "matching" mouse and Influenza strains. Otherwise, infection could vary from asymptomatic to atypically severe compared to human infections. Other animal models that are naturally susceptible to human Influenza strains such as hamsters, ferrets and guinea pigs could be utilized for these studies [117].

## 6 Conclusions

The multiple advantages, and potential pitfalls or problems posed by drug repurposing have been previously described [118]. Among the outstanding advantages are the low cost and reduced time for drug development, known pharmacological properties and suspected or confirmed interaction with pathogen targets determined through mechanistic studies [119].

The current review illustrates how drug repurposing efforts have led to the identification of several antihistamines as potential anti-infective drugs for a wide range of pathogen phyla. The majority are second generation antihistamines with lower side effects than first generation antihistamines. We observed that discoveries of drug repurposing relied mostly on *in vitro* studies occasionally supported by modest numbers of *in vivo* preclinical data. The notorious absence of clinical trials is somewhat surprising because these drugs have gone through safety studies for their original indication, and consequently they only require demonstrating efficacy. On the other hand, toxicity assessment may be a necessary step and a potential hurdle due to the eventual need to utilize larger doses than those approved to treat allergies. The difficulty in achieving significant toxicity improvement of some antihistamine analogues through medicinal chemistry can be overcome by means of combinatorial therapy as the most sensible and realistic approach. In fact, it is reasonable to assume that most of the antihistamines referenced in this review could efficiently complement current anti-infectious drugs, thereby reducing the development of drug resistance, toxicity and improving compliance.

The scarcity of clinical trials is still the most challenging aspect of drug repurposing. This could be explained by the low incentive for pharmaceutical companies to invest in costly clinical trials involving drugs that are off patent. To overcome these obstacles, increased collaborations of multidisciplinary research groups in academia with granting agencies and government are crucial to support clinical trials. Therefore, only through inter-institutional participation it will be possible to "harvest these low hanging fruits" and increase the arsenal of anti-infective drugs.

**Conflict of Interest**

The author confirms there is no known conflict of interest associated with this publication

**Acknowledgements**

This review article did not receive any specific grant from funding agencies in the public, commercial, or not-for-profit sectors.

**Bibliography**

1. Mercorelli B, Palu G, Loregian A. Drug Repurposing for Viral Infectious Diseases: How Far Are We? Trends Microbiol. 2018;26(10):865-76.
2. Zheng W, Sun W, Simeonov A. Drug repurposing screens and synergistic drug-combinations for infectious diseases. Br J Pharmacol. 2018;175(2):181-91.
3. Farha MA, Brown ED. Drug repurposing for antimicrobial discovery. Nature microbiology. 2019;4(4):565-77.
4. Fein MN, Fischer DA, O'Keefe AW, Sussman GL. CSACI position statement: Newer generation H1-antihistamines are safer than first-generation H1-antihistamines and should be the first-line antihistamines for the treatment of allergic rhinitis and urticaria. Allergy Asthma Clin Immunol. 2019;15:61.
5. Akdis CA, Simons FE. Histamine receptors are hot in immunopharmacology. Eur J Pharmacol. 2006;533(1-3):69-76.
6. Organization WH. World malaria report 2020: 20 years of global progress and challenges. 2020.
7. Dunst J, Kamena F, Matuschewski K. Cytokines and chemokines in cerebral malaria pathogenesis. Frontiers in cellular and infection microbiology. 2017;7:324.
8. Organization WH. Guidelines for the treatment of malaria: World Health Organization; 2015.
9. Hounkpatin AB, Kreidenweiss A, Held J. Clinical utility of tafenoquine in the prevention of relapse of Plasmodium vivax malaria: a review on the mode of action and emerging trial data. Infection and Drug Resistance. 2019;12:553.
10. Parhizgar AR, Tahghighi A. Introducing New Antimalarial Analogues of Chloroquine and Amodiaquine: A Narrative Review. Iran J Med Sci. 2017;42(2):115-28.
11. Adebayo JO, Tijjani H, Adegunloye AP, Ishola AA, Balogun EA, Malomo SO. Enhancing the antimalarial activity of artesunate. Parasitol Res. 2020;119(9):2749-64.
12. Haldar K, Bhattacharjee S, Safeukui I. Drug resistance in Plasmodium. Nat Rev Microbiol. 2018;16(3):156-70.
13. Chong CR, Chen X, Shi L, Liu JO, Sullivan DJ, Jr. A clinical drug library screen identifies astemizole as an antimalarial agent. Nat Chem Biol. 2006;2(8):415-6.
14. Lindquist M, Edwards IR. Risks of non-sedating antihistamines. Lancet. 1997;349(9061):1322.
15. Musonda CC, Whitlock GA, Witty MJ, Brun R, Kaiser M. Chloroquine-astemizole hybrids with potent in vitro and in vivo antiplasmodial activity. Bioorg Med Chem Lett. 2009;19(2):481-4.
16. Roman G, Crandall IE, Szarek WA. Synthesis and anti-Plasmodium activity of benzimidazole analogues structurally related to astemizole. ChemMedChem. 2013;8(11):1795-804.
17. Derbyshire ER, Prudencio M, Mota MM, Clardy J. Liver-stage malaria parasites vulnerable to diverse chemical scaffolds. Proceedings of the National Academy of Sciences of the United States of America. 2012;109(22):8511-6.
18. Roth A, Maher SP, Conway AJ, Ubalee R, Chaumeau V, Andolina C, et al. A comprehensive model for assessment of liver stage therapies targeting Plasmodium vivax and Plasmodium falciparum. Nature communications. 2018;9(1):1837.
19. Roth A, Maher SP, Conway AJ, Ubalee R, Chaumeau V, Andolina C, et al. Author Correction: A comprehensive model for assessment of liver stage therapies targeting Plasmodium vivax and Plasmodium falciparum. Nature communications. 2018;9(1):2317.
20. Voorberg-van der Wel A, Kocken CHM, Zeeman AM. Modeling Relapsing Malaria: Emerging Technologies to Study Parasite-Host Interactions in the Liver. Front Cell Infect Microbiol. 2020;10:606033.

21. Mikolajczak SA, Vaughan AM, Kangwanrangsan N, Roobsoong W, Fishbaugher M, Yimamnuaychok N, et al. Plasmodium vivax liver stage development and hypnozoite persistence in human liver-chimeric mice. Cell Host Microbe. 2015;17(4):526-35.
22. Kumar M, Okombo J, Mambwe D, Taylor D, Lawrence N, Reader J, et al. Multistage antiplasmodium activity of astemizole analogues and inhibition of hemozoin formation as a contributor to their mode of action. ACS infectious diseases. 2018;5(2):303-15.
23. Tian J, Vandermosten L, Peigneur S, Moreels L, Rozenski J, Tytgat J, et al. Astemizole analogues with reduced hERG inhibition as potent antimalarial compounds. Bioorg Med Chem. 2017;25(24):6332-44.
24. Sanguinetti MC, Tristani-Firouzi M. hERG potassium channels and cardiac arrhythmia. Nature. 2006;440(7083):463-9.
25. Giaginis C, Tsopelas F, Tsantili-Kakoulidou A. The impact of lipophilicity in drug discovery: rapid measurements by means of reversed-phase HPLC.  Rational Drug Design: Springer; 2018. p. 217-28.
26. Lu KY, Quan B, Sylvester K, Srivastava T, Fitzgerald MC, Derbyshire ER. Plasmodium chaperonin TRiC/CCT identified as a target of the antihistamine clemastine using parallel chemoproteomic strategy. Proceedings of the National Academy of Sciences of the United States of America. 2020;117(11):5810-7.
27. Kapishnikov S, Staalso T, Yang Y, Lee J, Perez-Berna AJ, Pereiro E, et al. Mode of action of quinoline antimalarial drugs in red blood cells infected by Plasmodium falciparum revealed in vivo. Proceedings of the National Academy of Sciences of the United States of America. 2019;116(46):22946-52.
28. WHO.https://www.who.int/health-topics/leishmaniasis. 2022.
29. Pradhan S, Schwartz R, Patil A, Grabbe S, Goldust M. Treatment options for leishmaniasis. Clinical and experimental dermatology. 2022;47(3):516-21.
30. Husein-ElAhmed H, Gieler U, Steinhoff M. Evidence supporting the enhanced efficacy of pentavalent antimonials with adjuvant therapy for cutaneous leishmaniasis: a systematic review and meta-analysis. Journal of the European Academy of Dermatology and Venereology. 2020;34(10):2216-28.
31. Chakraborty AK, Majumder HK. Mode of action of pentavalent antimonials: Specific inhibition of type I DNA topoisomerase of Leishmaniadonovani. Biochemical and biophysical research communications. 1988;152(2):605-11.
32. Walker J, Gongora R, Vasquez J-J, Drummelsmith J, Burchmore R, Roy G, et al. Discovery of factors linked to antimony resistance in Leishmania panamensis through differential proteome analysis. Mol Biochem Parasit. 2012;183(2):166-76.
33. Frezard F, Demicheli C, Kato KC, Reis PG, Lizarazo-Jaimes EH. Chemistry of antimony-based drugs in biological systems and studies of their mechanism of action. Reviews in Inorganic Chemistry. 2013;33(1):1-12.
34. Rajasekaran R, Chen Y-PP. Potential therapeutic targets and the role of technology in developing novel antileishmanial drugs. Drug discovery today. 2015;20(8):958-68.
35. Mishra J, Singh S. Miltefosine resistance in Leishmania donovani involves suppression of oxidative stress-induced programmed cell death. Experimental parasitology. 2013;135(2):397-406.
36. Saha AK, Mukherjee T, Bhaduri A. Mechanism of action of amphotericin B on Leishmania donovani promastigotes. Mol Biochem Parasit. 1986;19(3):195-200.
37. Braga SS. Multi-target drugs active against leishmaniasis: A paradigm of drug repurposing. European Journal of Medicinal Chemistry. 2019;183:111660.
38. Peniche AG, Osorio EY, Melby PC, Travi BL. Efficacy of histamine H1 receptor antagonists azelastine and fexofenadine against cutaneous Leishmania major infection. Plos Neglect Trop D. 2020;14(8):e0008482.

39. Pinto EG, da Costa-Silva TA, Tempone AG. Histamine H1-receptor antagonists against Leishmania (L.) infantum: an in vitro and in vivo evaluation using phosphatidylserine-liposomes. Acta tropica. 2014;137:206-10.
40. Reimão JQ, Scotti MT, Tempone AG. Anti-leishmanial and anti-trypanosomal activities of 1, 4-dihydropyridines: In vitro evaluation and structure–activity relationship study. Bioorganic & Medicinal Chemistry. 2010;18(22):8044-53.
41. de Melo Mendes V, Tempone AG, Borborema SET. Antileishmanial activity of H1-antihistamine drugs and cellular alterations in Leishmania (L.) infantum. Acta tropica. 2019;195:6-14.
42. Mina JG, Charlton RL, Alpizar-Sosa E, Escrivani DO, Brown C, Alqaisi A, et al. Antileishmanial chemotherapy through clemastine fumarate mediated inhibition of the leishmania inositol phosphorylceramide synthase. ACS infectious diseases. 2020;7(1):47-63.
43. Drugbank.https://go.drugbank.com/drugs/DB00283. 2022.
44. López-Vélez R, Norman FF, Bern C. American trypanosomiasis (Chagas disease). Hunter's Tropical Medicine and Emerging Infectious Diseases: Elsevier; 2020. p. 762-75.
45. Rajao MA, Furtado C, Alves CL, Passos-Silva DG, de Moura MB, Schamber-Reis BL, et al. Unveiling benznidazole's mechanism of action through overexpression of DNA repair proteins in Trypanosoma cruzi. Environmental and molecular mutagenesis. 2014;55(4):309-21.
46. Hof H. A new, broad-spectrum azole antifungal: posaconazole–mechanisms of action and resistance, spectrum of activity. Mycoses. 2006;49:2-6.
47. Drugbank.https://go.drugbank.com/drugs/DB11820. 2022.
48. Planer JD, Hulverson MA, Arif JA, Ranade RM, Don R, Buckner FS. Synergy testing of FDA-approved drugs identifies potent drug combinations against Trypanosoma cruzi. Plos Neglect Trop D. 2014;8(7):e2977.
49. Booth L-A, Smith TK. Lipid metabolism in Trypanosoma cruzi: A review. Mol Biochem Parasit. 2020;240:111324.
50. Urbina JA. Recent clinical trials for the etiological treatment of chronic Chagas disease: advances, challenges and perspectives. Journal of Eukaryotic Microbiology. 2015;62(1):149-56.
51. Chatelain E, Konar N. Translational challenges of animal models in Chagas disease drug development: a review. Drug Des Devel Ther 9: 4807–4823. 2015.
52. Meyers AC, Hamer SA, Matthews D, Gordon SG, Saunders AB. Risk factors and select cardiac characteristics in dogs naturally infected with Trypanosoma cruzi presenting to a teaching hospital in Texas. Journal of veterinary internal medicine. 2019;33(4):1695-706.
53. Gryseels B, Polman K, Clerinx J, Kestens L. Human schistosomiasis. The Lancet. 2006;368(9541):1106-18.
54. Vale N, Gouveia MJ, Rinaldi G, Brindley PJ, Gärtner F, Correia da Costa JM. Praziquantel for schistosomiasis: single-drug metabolism revisited, mode of action, and resistance. Antimicrobial agents and chemotherapy. 2017;61(5):e02582-16.
55. Fallon PG, Doenhoff MJ. Drug-resistant schistosomiasis: resistance to praziquantel and oxamniquine induced in Schistosoma mansoni in mice is drug specific. The American journal of tropical medicine and hygiene. 1994;51(1):83-8.
56. Sadhu PS, Kumar SN, Chandrasekharam M, Pica-Mattoccia L, Cioli D, Rao VJ. Synthesis of new praziquantel analogues: potential candidates for the treatment of schistosomiasis. Bioorganic & medicinal chemistry letters. 2012;22(2):1103-6.
57. Roquini DB, Cogo RM, Mengarda AC, Mazloum SF, Morais CS, Xavier RP, et al. Promethazine exhibits antiparasitic properties in vitro and reduces worm burden, egg production, hepatomegaly, and splenomegaly in a schistosomiasis animal model. Antimicrobial Agents and Chemotherapy. 2019;63(12):e01208-19.

58. Kimber MJ, Sayegh L, El-Shehabi F, Song C, Zamanian M, Woods DJ, et al. Identification of an Ascaris G protein-coupled acetylcholine receptor with atypical muscarinic pharmacology. International journal for parasitology. 2009;39(11):1215-22.
59. MacDonald K, Kimber MJ, Day TA, Ribeiro P. A constitutively active G protein-coupled acetylcholine receptor regulates motility of larval Schistosoma mansoni. Mol Biochem Parasit. 2015;202(1):29-37.
60. Zhao Y, Lin J, Fan Y, Lin X. Life cycle of Cryptococcus neoformans. Annual review of microbiology. 2019;73:17-42.
61. Bermas A, Geddes-McAlister J. Combatting the evolution of antifungal resistance in Cryptococcus neoformans. Molecular microbiology. 2020;114(5):721-34.
62. Jirí Houšt JS, Havlícek V. Antifungal Drugs.
63. Iyer KR, Revie NM, Fu C, Robbins N, Cowen LE. Treatment strategies for cryptococcal infection: challenges, advances and future outlook. Nature Reviews Microbiology. 2021;19(7):454-66.
64. Hole CR, Wormley Jr FL. Vaccine and immunotherapeutic approaches for the prevention of cryptococcosis: lessons learned from animal models. Frontiers in microbiology. 2012;3:291.
65. Vu K, Gelli A. Astemizole and an analogue promote fungicidal activity of fluconazole against Cryptococcus neoformans var. grubii and Cryptococcus gattii. Medical Mycology. 2010;48(2):255-62.
66. Pereira TC, De Barros PP, Fugisaki LRdO, Rossoni RD, Ribeiro FdC, De Menezes RT, et al. Recent advances in the use of Galleria mellonella model to study immune responses against human pathogens. Journal of Fungi. 2018;4(4):128.
67. Lu R, Hollingsworth C, Qiu J, Wang A, Hughes E, Xin X, et al. Efficacy of oral encochleated amphotericin B in a mouse model of cryptococcal meningoencephalitis. mBio. 2019;10(3):e00724-19.
68. Barton M, O'Brien K, Robinson JL, Davies DH, Simpson K, Asztalos E, et al. Invasive candidiasis in low birth weight preterm infants: risk factors, clinical course and outcome in a prospective multicenter study of cases and their matched controls. BMC infectious diseases. 2014;14(1):1-10.
69. Greenberg RG, Benjamin Jr DK. Neonatal candidiasis: diagnosis, prevention, and treatment. Journal of Infection. 2014;69:S19-S22.
70. Logan C, Martin-Loeches I, Bicanic T. Invasive candidiasis in critical care: challenges and future directions. Intensive Care Medicine. 2020;46(11):2001-14.
71. Ben-Ami R. Treatment of invasive candidiasis: A narrative review. Journal of Fungi. 2018;4(3):97.
72. Dennis EK, Garneau-Tsodikova S. Synergistic combinations of azoles and antihistamines against Candida species in vitro. Medical Mycology. 2019;57(7):874-84.
73. Miesel L, Lin KY, Ong V. Rezafungin treatment in mouse models of invasive candidiasis and aspergillosis: Insights on the PK/PD pharmacometrics of rezafungin efficacy. Pharmacology Research & Perspectives. 2019;7(6):e00546.
74. Hirayama T, Miyazaki T, Ito Y, Wakayama M, Shibuya K, Yamashita K, et al. Virulence assessment of six major pathogenic Candida species in the mouse model of invasive candidiasis caused by fungal translocation. Scientific reports. 2020;10(1):1-10.
75. Weiner L, Webb A, Limbago B. Antimicrobial-Resistant Pathogens 366 Associated With Healthcare-Associated Infections: Summary of Data Reported to 367 the National Healthcare Safety Network at the Centers for Disease Control and 368 Prevention, 2011–2014. Infect Control Hosp Epidemiol. 2016.
76. Van Duin D, Paterson DL. Multidrug-resistant bacteria in the community: trends and lessons learned. Infectious disease clinics. 2016;30(2):377-90.
77. Liu C, Bayer A, Cosgrove SE, Daum RS, Fridkin SK, Gorwitz RJ, et al. Clinical practice guidelines by the Infectious Diseases Society of America for the treatment of methicillin-resistant Staphylococcus aureus infections in adults and children. Clinical infectious diseases. 2011;52(3):e18-e55.

78. Boneca IG, Chiosis G. Vancomycin resistance: occurrence, mechanisms and strategies to combat it. Expert opinion on therapeutic targets. 2003;7(3):311-28.
79. Perlmutter JI, Forbes LT, Krysan DJ, Ebsworth-Mojica K, Colquhoun JM, Wang JL, et al. Repurposing the antihistamine terfenadine for antimicrobial activity against Staphylococcus aureus. Journal of medicinal chemistry. 2014;57(20):8540-62.
80. Waring MJ. Lipophilicity in drug discovery. Expert Opinion on Drug Discovery. 2010;5(3):235-48.
81. Cutrona N, Gillard K, Ulrich R, Seemann M, Miller HB, Blackledge MS. From antihistamine to anti-infective: loratadine inhibition of regulatory PASTA kinases in Staphylococci reduces biofilm formation and potentiates β-lactam antibiotics and vancomycin in resistant strains of Staphylococcus aureus. ACS infectious diseases. 2019;5(8):1397-410.
82. Henz B. The pharmacologic profile of desloratadine: a review. Allergy. 2001;56:7-13.
83. Simons F. H1-receptor antagonists. Drug safety. 1994;10(5):350-80.
84. Vergidis P, Rouse MS, Euba G, Karau MJ, Schmidt SM, Mandrekar JN, et al. Treatment with linezolid or vancomycin in combination with rifampin is effective in an animal model of methicillin-resistant Staphylococcus aureus foreign body osteomyelitis. Antimicrobial agents and chemotherapy. 2011;55(3):1182-6.
85. Jia M, Chen Z, Du X, Guo Y, Sun T, Zhao X. A simple animal model of Staphylococcus aureus biofilm in sinusitis. American Journal of Rhinology & Allergy. 2014;28(2):e115-e9.
86. Kausar S, Said Khan F, Ishaq Mujeeb Ur Rehman M, Akram M, Riaz M, Rasool G, et al. A review: Mechanism of action of antiviral drugs. International Journal of Immunopathology and Pharmacology. 2021;35:20587384211002621.
87. NIH. Overview of COVID-19. https://filescovid19treatmentguidelinesnihgov/guidelines/section/section_9pdf. 2022.
88. Zhang J-j, Dong X, Liu G-h, Gao Y-d. Risk and Protective Factors for COVID-19 Morbidity, Severity, and Mortality. Clinical Reviews in Allergy & Immunology. 2022:1-18.
89. Wen W, Chen C, Tang J, Wang C, Zhou M, Cheng Y, et al. Efficacy and safety of three new oral antiviral treatment (molnupiravir, fluvoxamine and Paxlovid) for COVID-19： a meta-analysis. Annals of Medicine. 2022;54(1):516-23.
90. Ge S, Lu J, Hou Y, Lv Y, Wang C, He H. Azelastine inhibits viropexis of SARS-CoV-2 spike pseudovirus by binding to SARS-CoV-2 entry receptor ACE2. Virology. 2021;560:110-5.
91. Hou Y, Ge S, Li X, Wang C, He H, He L. Testing of the inhibitory effects of loratadine and desloratadine on SARS-CoV-2 spike pseudotyped virus viropexis. Chemico-biological interactions. 2021;338:109420.
92. Ge S, Wang X, Hou Y, Lv Y, Wang C, He H. Repositioning of histamine H1 receptor antagonist: Doxepin inhibits viropexis of SARS-CoV-2 Spike pseudovirus by blocking ACE2. European Journal of Pharmacology. 2021;896:173897.
93. Yang L, Pei R-j, Li H, Ma X-n, Zhou Y, Zhu F-h, et al. Identification of SARS-CoV-2 entry inhibitors among already approved drugs. Acta Pharmacologica Sinica. 2021;42(8):1347-53.
94. Reznikov LR, Norris MH, Vashisht R, Bluhm AP, Li D, Liao Y-SJ, et al. Identification of antiviral antihistamines for COVID-19 repurposing. Biochemical and biophysical research communications. 2021;538:173-9.
95. Pandamooz S, Jurek B, Meinung C-P, Baharvand Z, Sahebi Shahem-abadi A, Haerteis S, et al. Experimental models of SARS-CoV-2 infection: possible platforms to study COVID-19 pathogenesis and potential treatments. Annual review of pharmacology and toxicology. 2022;62:25-53.
96. Rugarabamu S, Mboera L, Rweyemamu M, Mwanyika G, Lutwama J, Paweska J, et al. Forty-two years of responding to Ebola virus outbreaks in Sub-Saharan Africa: a review. BMJ Global Health. 2020;5(3):e001955.

97. Asad A, Aamir A, Qureshi NE, Bhimani S, Jatoi NN, Batra S, et al. Past and current advances in Marburg virus disease: a review. Le Infezioni in Medicina. 2020;28(3):332-45.
98. Iversen PL, Kane CD, Zeng X, Panchal RG, Warren TK, Radoshitzky SR, et al. Recent successes in therapeutics for Ebola virus disease: no time for complacency. The Lancet Infectious Diseases. 2020;20(9):e231-e7.
99. Schafer A, Cheng H, Xiong R, Soloveva V, Retterer C, Mo F, et al. Repurposing potential of 1st generation H1-specific antihistamines as anti-filovirus therapeutics. Antiviral research. 2018;157:47-56.
100. Cross RW, Fenton KA, Geisbert TW. Small animal models of filovirus disease: recent advances and future directions. Expert Opinion on Drug Discovery. 2018;13(11):1027-40.
101. Jafri SM, Gordon SC. Epidemiology of hepatitis C. Clinical liver disease. 2018;12(5):140.
102. Goto K, Roca Suarez AA, Wrensch F, Baumert TF, Lupberger J. Hepatitis C virus and hepatocellular carcinoma: when the host loses its grip. International journal of molecular sciences. 2020;21(9):3057.
103. Eltahla AA, Luciani F, White PA, Lloyd AR, Bull RA. Inhibitors of the hepatitis C virus polymerase; mode of action and resistance. Viruses. 2015;7(10):5206-24.
104. Baumert TF, Berg T, Lim JK, Nelson DR. Status of direct-acting antiviral therapy for hepatitis C virus infection and remaining challenges. Gastroenterology. 2019;156(2):431-45.
105. Sorbo MC, Cento V, Di Maio VC, Howe AY, Garcia F, Perno CF, et al. Hepatitis C virus drug resistance associated substitutions and their clinical relevance: Update 2018. Drug Resistance Updates. 2018;37:17-39.
106. He S, Lin B, Chu V, Hu Z, Hu X, Xiao J, et al. Repurposing of the antihistamine chlorcyclizine and related compounds for treatment of hepatitis C virus infection. Science translational medicine. 2015;7(282):282ra49-ra49.
107. Hu Z, Rolt A, Hu X, Ma CD, Le DJ, Park SB, et al. Chlorcyclizine inhibits viral fusion of hepatitis C virus entry by directly targeting HCV envelope glycoprotein 1. Cell chemical biology. 2020;27(7):780-92. e5.
108. Rolt A, Le D, Hu Z, Wang AQ, Shah P, Singleton M, et al. Preclinical pharmacological development of chlorcyclizine derivatives for the treatment of hepatitis C virus infection. The Journal of infectious diseases. 2018;217(11):1761-9.
109. Koh C, Dubey P, Han MAT, Walter PJ, Garraffo HM, Surana P, et al. A randomized, proof-of-concept clinical trial on repurposing chlorcyclizine for the treatment of chronic hepatitis C. Antiviral research. 2019;163:149-55.
110. Poluzzi E, Raschi E, Godman B, Koci A, Moretti U, Kalaba M, et al. Pro-arrhythmic potential of oral antihistamines (H1): combining adverse event reports with drug utilization data across Europe. PLoS One. 2015;10(3):e0119551.
111. Salama G, London B. Mouse models of long QT syndrome. The Journal of physiology. 2007;578(1):43-53.
112. Kettenhofen R, Bohlen H. Preclinical assessment of cardiac toxicity. Drug discovery today. 2008;13(15-16):702-7.
113. Mauskopf J, Klesse M, Lee S, Herrera-Taracena G. The burden of influenza complications in different high-risk groups: a targeted literature review. Journal of medical economics. 2013;16(2):264-77.
114. Wang X, Li Y, O'Brien KL, Madhi SA, Widdowson M-A, Byass P, et al. Global burden of respiratory infections associated with seasonal influenza in children under 5 years in 2018: a systematic review and modelling study. The Lancet Global Health. 2020;8(4):e497-e510.
115. Principi N, Camilloni B, Alunno A, Polinori I, Argentiero A, Esposito S. Drugs for influenza treatment: is there significant news? Frontiers in medicine. 2019;6:109.

116.     Xu W, Xia S, Pu J, Wang Q, Li P, Lu L, et al. The antihistamine drugs carbinoxamine maleate and chlorpheniramine maleate exhibit potent antiviral activity against a broad spectrum of influenza viruses. Frontiers in microbiology. 2018:2643.
117.     Bouvier NM, Lowen AC. Animal models for influenza virus pathogenesis and transmission. Viruses. 2010;2(8):1530-63.
118.     Mercorelli B, Palù G, Loregian A. Drug repurposing for viral infectious diseases: how far are we? Trends in microbiology. 2018;26(10):865-76.
119.     Farha MA, Brown ED. Drug repurposing for antimicrobial discovery. Nature microbiology. 2019;4(4):565-77.